\RequirePackage{fix-cm}
\documentclass[twocolumn,epjc3]{svjour3}  
\usepackage{graphicx} % Required for inserting images
\usepackage{amsmath,amssymb}
\usepackage{booktabs}
\smartqed  % flush right qed marks, e.g. at end of proof
\RequirePackage{graphicx}

\journalname{Eur. Phys. J. C}
\begin{document}

\title{Orbital Dynamics and Gravitational-Wave Signatures of EMRIs in Self-Dual Loop Quantum Gravity Black Holes}
\author{Yu Wang\thanksref{e1,addr1,addr2}\and Meilin Liu\thanksref{e2,addr2}
\and Haiguang Xu\thanksref{addr1}}
\thankstext{e1}{e-mail: sjtu2686361@sjtu.edu.cn (Yu Wang)}
\thankstext{e2}{Corresponding author: meilin.liu@sjtu.edu.cn (Meilin Liu)}

\institute{School of Physics and Astronomy, Shanghai Jiao Tong University, Shanghai 200240, China \label{addr1}
           \and
           School of Aeronautics and Astronautics, Shanghai Jiao Tong University, Shanghai 200240, China \label{addr2}
}
\date{\today}
\maketitle

\begin{abstract}

Loop quantum gravity (LQG) predicts quantum modifications to classical black-hole spacetimes, which may leave imprints on the dynamics and gravitational-wave signals of compact objects in the strong-field regime. In this work, we investigate the orbital dynamics and gravitational-wave signatures of extreme mass-ratio inspirals (EMRIs) in a self-dual loop quantum gravity black hole spacetime. We analyze test-particle motion in the static, spherically symmetric self-dual LQG geometry characterized by two quantum parameters: the polymeric parameter $P$ and the minimal area parameter $a_0$. The effective potential and orbital structure are systematically studied, and we quantify the influence of quantum corrections on circular-orbit stability and strong-field orbital behavior. Compared with the classical Schwarzschild spacetime, LQG corrections modify the near-horizon orbital dynamics. Based on the orbital evolution, we construct gravitational-wave waveforms and investigate the impact of quantum corrections on waveform morphology. We find that LQG effects accumulate during the long inspiral phase, leading to noticeable signal deviations from the classical case. To incorporate rotation, we construct a rotating extension of the self-dual spacetime using the Newman--Janis algorithm. The resulting LQG-corrected Kerr geometry is used to analyze orbital motion, revealing the interplay between spin and quantum corrections in strong-field trajectories. Finally, we perform a Fisher matrix analysis to estimate potential constraints on quantum parameters from future space-based gravitational-wave detectors. Our results indicate that EMRI observations provide a promising avenue to probe quantum gravitational effects in black-hole spacetimes.


\end{abstract}

\section{Introduction}

The detection of gravitational waves has opened a new observational window into the strong-field regime of gravity, providing unprecedented opportunities to test general relativity in the dynamical and nonlinear regime \cite{Abbott2016,Abbott2019}. Among the most promising sources for space-based detectors such as LISA are extreme mass-ratio inspirals (EMRIs), where a stellar-mass compact object orbits a massive black hole over long timescales \cite{AmaroSeoane2017}. Due to the large number of orbital cycles accumulated in the strong-field region, EMRIs are extremely sensitive to small deviations from the Kerr geometry, making them powerful probes of fundamental physics beyond general relativity \cite{BarackCutler2004,Gair2013,Barack2018}.Accurate modeling of EMRIs requires a precise understanding of geodesic motion and radiation reaction effects in curved spacetime \cite{Pound2015,CutlerFlanagan1994}. In particular, small modifications of the background geometry may accumulate over long inspiral times and lead to observable deviations in gravitational-wave signals, making EMRIs an ideal probe of new gravitational physics \cite{Maselli2020,Stein2021}.

Loop quantum gravity (LQG) provides a non-perturbative approach to quantum gravity in which spacetime geometry becomes discrete at the Planck scale \cite{Rovelli2004,Ashtekar2004,Thiemann2007}. Within effective semiclassical descriptions, LQG leads to quantum-corrected black-hole geometries that resolve classical singularities while preserving the correct infrared limit. To date, various effective black hole spacetimes derived from LQG have been constructed, incorporating the leading quantum corrections while maintaining agreement with classical general relativity in the infrared limit \cite{Modesto2004,Modesto2006,SelfDual2009,Modesto2010,Gambini2013}. Parallel to these developments, the canonical quantization procedure for spherically symmetric LQG black holes has been firmly established, along with detailed studies of the Hamiltonian constraint, volume operator, and quantum horizon structures within this reduced framework \cite{Thiemann1993,Campiglia2007,Bengtsson1988,Bojowald2000,Bojowald2004,Bojowald2005,Bojowald2006,Kuchar1994}.

Various loop-quantum-gravity-inspired black-hole models have been proposed and extensively studied in recent years \cite{Perez2017,Bojowald2020}.In particular, self-dual loop quantum gravity black holes provide a phenomenological framework incorporating key quantum corrections characterized by two parameters, which modify the near-horizon geometry while reducing to the Schwarzschild solution in the classical limit. The orbital structure and geodesic motion in such spacetimes have been investigated in recent literature, revealing significant deviations from classical behavior in the strong-field regime \cite{Tu2023,ISCO2024LQG,SelfDualGeodesics2024,ZiKumar2025}.From an observational perspective, quantum-gravity-induced modifications of black-hole spacetimes may lead to measurable effects in astrophysical phenomena such as black-hole shadows, gravitational lensing, and quasi-periodic oscillations \cite{LQGShadow2023,QPO2023}. These studies suggest that high-precision observations could provide constraints on deviations from general relativity in the strong-field regime.

In this work, we investigate the orbital dynamics of test particles in a self-dual loop quantum gravity black hole spacetime characterized by two quantum parameters. We analyze the effective potential and orbital structure, focusing on the impact of quantum corrections on circular orbit stability and strong-field motion.Based on the geodesic dynamics, we construct the corresponding gravitational-wave waveforms associated with EMRIs and study the influence of quantum corrections on the emitted signals. The waveforms are consistently generated from orbital motion, allowing cumulative effects in the strong-field regime to be captured.To incorporate astrophysically realistic rotation, we construct a rotating extension of the self-dual spacetime using the Newman--Janis algorithm \cite{NewmanJanis1965,Erbin2017}. The resulting LQG-corrected Kerr geometry allows us to study the interplay between spin and quantum corrections in determining orbital dynamics \cite{RotatingLQG2024}.Finally, we perform a Fisher matrix analysis to estimate the potential constraints on the quantum parameters from future space-based gravitational-wave observations. Our results suggest that EMRI systems provide a promising avenue for probing quantum gravitational effects in black-hole spacetimes.

The structure of this paper is as follows. In Sec.~2, we introduce the self-dual LQG black hole spacetime and analyze its effective potential. In Sec.~3, we study the orbital dynamics and gravitational-wave waveforms. In Sec.~4, we construct the rotating extension via the Newman--Janis algorithm. In Sec.~5, we present the Fisher matrix analysis. Finally, Sec.~6 concludes the paper.

\section{Loop Quantum Gravity Corrected Black Hole Spacetime}

In this section, we introduce the loop quantum gravity (LQG) inspired effective black hole geometry adopted in the present work. Various effective black hole solutions motivated by loop quantum gravity have been proposed in the literature, typically arising from polymerization techniques or semiclassical effective treatments of quantum geometry\cite{Cruz2019}. These models generally introduce correction parameters that modify the near-horizon structure while preserving the classical Schwarzschild limit at large distances.

We consider a static and spherically symmetric LQG-corrected spacetime described by the line element

\begin{equation}
ds^2 = -G(r)\,dt^2 + \frac{dr^2}{F(r)} + H(r)\left(d\theta^2 + \sin^2\theta\,d\phi^2\right),
\end{equation}

where the metric functions are given by

\begin{align}
G(r) &= \frac{(r-r_p)(r-r_m)(r+r_\ast)^2}{r^4 + a_0^2}, \\
F(r) &= \frac{(r-r_p)(r-r_m)r^4}{(r+r_\ast)^2 (r^4 + a_0^2)}, \\
H(r) &= r^2 + \frac{a_0^2}{r^2}.
\end{align}

Here the parameters are defined as

\begin{equation}
r_p = 2M, \quad r_m = 2MP^2, \quad r_\ast = 2MP,
\end{equation}

where $M$ is the black hole mass, while $P$ and $a_0$ characterize quantum gravity corrections.

In the classical limit $P \rightarrow 0$ and $a_0 \rightarrow 0$, the metric reduces to the Schwarzschild spacetime,

\begin{equation}
G(r) = F(r) = 1 - \frac{2M}{r}, \qquad H(r)=r^2.
\end{equation}

The correction parameter $a_0$ modifies the strong-field region and removes the classical curvature singularity at $r \to 0$, while $P$ controls the deviation from the classical horizon structure. These quantum corrections become significant in the near-horizon regime and affect both geodesic motion and gravitational wave emission from compact object systems.

The event horizon is determined by the condition

\begin{equation}
G(r_h)=0,
\end{equation}

which generally deviates from the Schwarzschild value $r_h = 2M$ due to quantum corrections.

Compared with the Schwarzschild case, the LQG correction modifies the horizon structure and the effective geometry experienced by orbiting particles. In particular, the near-horizon region becomes regularized, reflecting the underlying quantum geometry effects.

To investigate the motion of a test particle of mass $\mu$, we consider the Lagrangian

\begin{equation}
\mathcal{L}=\frac{1}{2}g_{\mu\nu}\dot{x}^{\mu}\dot{x}^{\nu},
\end{equation}

where the overdot denotes differentiation with respect to the proper time $\tau$.

Due to stationarity and spherical symmetry, the spacetime admits two conserved quantities: the energy $E$ and the angular momentum $L$, given by

\begin{equation}
E = G(r)\,\dot{t},
\end{equation}
\begin{equation}
L = H(r)\,\dot{\phi}.
\end{equation}

Restricting the motion to the equatorial plane $\theta=\pi/2$, the radial equation of motion can be written as
\begin{equation}
\dot{r}^2 + V_{\rm eff}(r) = E^2,
\end{equation}
where the effective potential is

\begin{equation}
V_{\rm eff}(r)
=
\frac{(r-r_p)(r-r_m)(r+r_\ast)^2}{r^4+a_0^2}
\left(
1+\frac{L^2 r^2}{r^4+a_0^2}
\right).
\end{equation}

The structure of the effective potential plays a crucial role in determining orbital dynamics and stability properties of the inspiraling compact object. In particular, the innermost stable circular orbit (ISCO) is determined by
\begin{equation}
\frac{dV_{\rm eff}}{dr}=0,
\end{equation}
and
\begin{equation}
\frac{d^2V_{\rm eff}}{dr^2}=0.
\end{equation}

Since EMRI gravitational waves are highly sensitive to orbital evolution in the strong-field region, even small shifts in the ISCO radius and orbital frequency may accumulate into significant gravitational-wave phase corrections over long inspiral timescales. Consequently, the quantum correction parameters $a_0$ and $P$ may leave observable signatures in both inspiral waveforms and black hole ringdown spectra.

In the following subsection, we investigate the effects of the LQG correction parameters on the effective potential, geodesic dynamics, and EMRI gravitational-wave waveforms in this spacetime.

\subsection{Effective potential and parameter dependence}

To further understand the influence of loop quantum gravity (LQG) corrections on orbital dynamics, we analyze the effective potential governing test-particle motion in the self-dual LQG black-hole spacetime. The effective potential is determined by the modified metric functions and depends on two quantum parameters: the polymeric parameter $P$ and the minimal area parameter $a_0$.

\begin{figure}[htbp]
    \centering
    \includegraphics[width=0.45\textwidth]{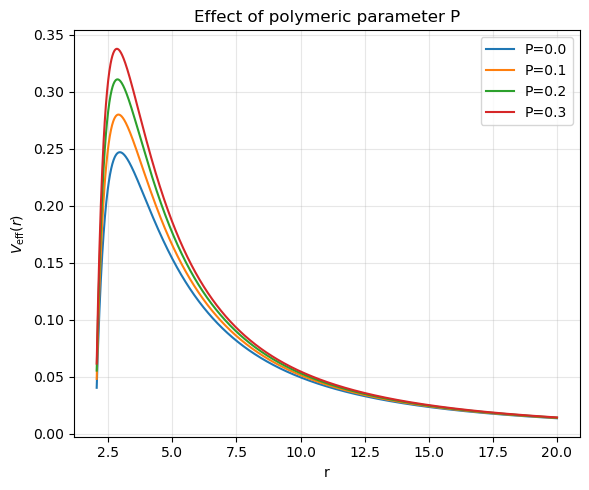}
    \caption{
    Effective potential $V_{\mathrm{eff}}(r)$ in the self-dual loop quantum gravity spacetime as a function of the polymeric parameter $P$, with fixed $a_0 = 0.05$. Increasing $P$ induces a global deformation of the potential barrier, shifting both the peak position and height.
    }
    \label{fig:potential_P}
\end{figure}

Figure~\ref{fig:potential_P} shows the effective potential for different values of $P$ with the minimal area parameter fixed at $a_0=0.05$. As $P$ increases, the height of the potential barrier decreases while its width becomes broader, whereas the peak position changes only slightly. This behavior indicates that the LQG corrections mainly smooth the effective gravitational potential in the near-horizon region, which may influence the orbital stability and inspiral dynamics of EMRI systems.

\begin{figure}[htbp]
    \centering
    \includegraphics[width=0.45\textwidth]{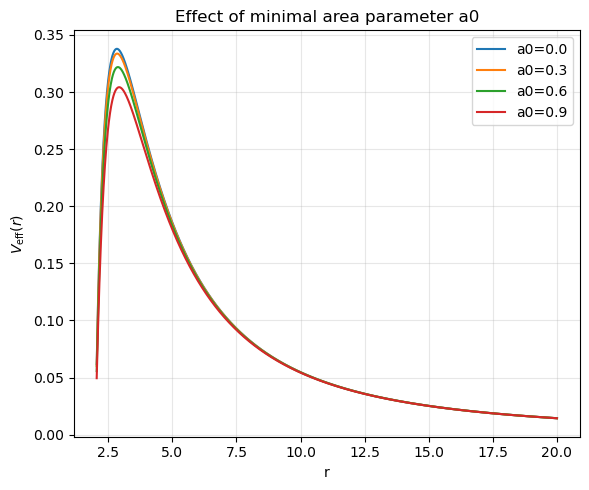}
    \caption{
    Effective potential $V_{\mathrm{eff}}(r)$ in the self-dual loop quantum gravity spacetime as a function of the minimal area parameter $a_0$, with fixed $P = 0.3$. The parameter $a_0$ mainly affects the small-radius regime, regularizing the near-singularity region and modifying the steepness of the potential barrier.
    }
    \label{fig:potential_a0}
\end{figure}

Figure~\ref{fig:potential_a0} shows the dependence of the effective potential $V_{\mathrm{eff}}(r)$ on the minimal area parameter $a_0$, while keeping $P=0.3$ fixed. In this case, variations in $a_0$ predominantly affect the small-radius region, where quantum geometry effects become significant. In particular, a nonzero $a_0$ regularizes the near-singularity geometry and modifies the shape of the effective potential barrier in the near-horizon region. These quantum corrections may influence the orbital dynamics of test particles and consequently affect the inspiral evolution of EMRI systems.

These results indicate that LQG corrections introduce scale-dependent modifications to the effective gravitational potential. The polymeric parameter $P$ controls the global structure of the potential barrier, whereas $a_0$ governs the near-horizon quantum geometry effects. Such features may lead to observable imprints in wave propagation and quasi-normal mode spectra.

\section{Orbit Dynamics and Gravitational-Wave Evolution}

In this section, we investigate the orbital dynamics of a compact object inspiraling into a loop quantum gravity (LQG)-corrected black hole and analyze the corresponding gravitational-wave evolution. Since extreme mass-ratio inspirals (EMRIs) undergo long-term evolution in the strong-field region around the central black hole, even small deviations from the classical spacetime geometry may accumulate over the inspiral process and leave observable imprints on the orbital dynamics and gravitational-wave signals.

We consider a compact object of mass $\mu$ orbiting a supermassive black hole of mass $M$, with the mass ratio satisfying
\begin{equation}
\mu \ll M .
\end{equation}
Under the test-particle approximation, the orbital motion is governed by the geodesic equations derived from the LQG-corrected metric introduced in the previous section.

For circular equatorial orbits, the orbital angular velocity is defined as
\begin{equation}
\Omega=\frac{d\phi}{dt}.
\end{equation}

Using the conserved quantities associated with the Killing symmetries of the spacetime, the angular velocity for a general static spherically symmetric spacetime can be written as
\begin{equation}
\Omega^2=\frac{G'(r)}{H'(r)},
\label{omega}
\end{equation}
where the prime denotes differentiation with respect to $r$. For the classical Schwarzschild spacetime, where $G(r)=1-2M/r$ and $H(r)=r^2$, this expression reduces to
\begin{equation}
\Omega_{\rm Sch}=\sqrt{\frac{M}{r^3}}.
\end{equation}

The LQG corrections modify the metric functions $G(r)$ and $H(r)$, leading to deviations in the orbital frequency and circular orbit structure, particularly in the strong-field region. As the compact object evolves toward the innermost stable circular orbit (ISCO), these quantum corrections can influence the orbital dynamics and the resulting gravitational-wave signals emitted during the inspiral process.

The specific energy and angular momentum of a particle moving on a circular orbit can be obtained from the conserved quantities associated with the time-translation and rotational symmetries. For the general LQG-corrected metric with the angular function $H(r)$, they are given by
\begin{equation}
E=\frac{G(r)}
{\sqrt{G(r)-H(r)\Omega^2}},
\end{equation}
and
\begin{equation}
L=\frac{H(r)\Omega}
{\sqrt{G(r)-H(r)\Omega^2}} .
\end{equation}

The inspiral evolution is driven by gravitational-wave emission from the compact binary system. Within the leading-order quadrupole approximation, the gravitational-wave energy flux is expressed as
\begin{equation}
\frac{dE}{dt}
=-\frac{32}{5}\mu^2 r^4\Omega^6 .
\label{flux}
\end{equation}

Combining the energy balance relation with the orbital energy of the circular orbit, the radial evolution of the inspiral can be obtained as
\begin{equation}
\frac{dr}{dt}
=
\frac{dE/dt}{dE/dr}.
\end{equation}

\subsection{Orbital dynamics in the self-dual LQG spacetime}

To investigate the impact of loop quantum gravity (LQG) corrections on test-particle dynamics, we compare geodesic motion in the Schwarzschild limit (GR) and in the effective self-dual LQG spacetime. The orbital dynamics is obtained by numerically integrating the Hamiltonian equations derived from the inverse metric $g^{\mu\nu}$.

We consider identical initial conditions for both systems, ensuring that the differences between the two trajectories arise solely from the quantum-gravity-induced modifications of the spacetime geometry. In particular, the LQG correction parameters modify the effective potential and influence the orbital evolution in the strong-field regime.

\begin{figure}[htbp]
    \centering
    \includegraphics[width=0.45\textwidth]{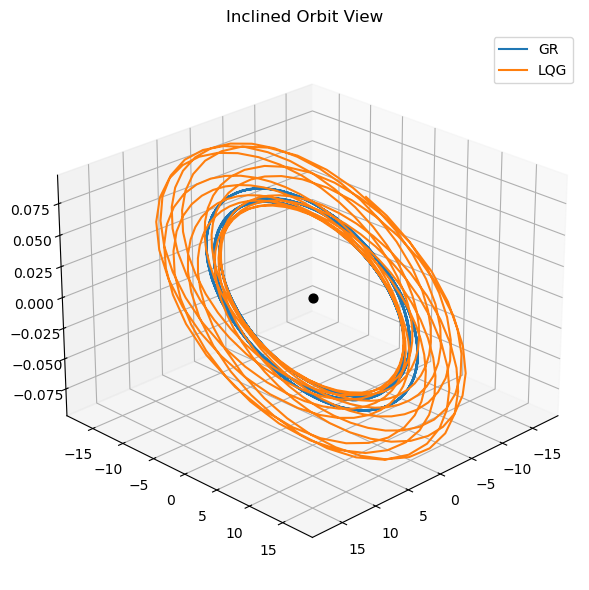}
   \caption{
Inclined-view comparison of geodesic motion in the Schwarzschild (GR) and self-dual LQG spacetimes.
The blue curve corresponds to the GR limit ($a_0=0$, $P=0$), while the orange curve represents the LQG-corrected orbit with quantum-gravity parameters ($a_0\neq0$, $P\neq0$).
Although the two trajectories overlap closely at early times, small deviations gradually accumulate during the evolution, leading to noticeable differences in the orbital precession and dynamical behavior.
The viewing angle $(\mathrm{elev}=25^\circ, \mathrm{azim}=45^\circ)$ highlights the deviations in the polar direction.
}
    \label{fig:orbit_inclined}
\end{figure}

Figure~\ref{fig:orbit_inclined} shows the resulting three-dimensional trajectories from an inclined view. The Schwarzschild case (GR) is represented by the blue curve, while the LQG-corrected orbit is shown by the orange curve. The deviation between the two trajectories reflects the influence of quantum corrections on the strong-field orbital dynamics.

These deviations originate from the modified metric components, which alter the effective radial dynamics and consequently affect the orbital evolution. Such effects, although perturbatively small, can accumulate over long inspiral timescales and may become relevant for extreme mass-ratio inspirals (EMRIs) observed by LISA.

To investigate the impact of loop quantum gravity (LQG) corrections on the orbital dynamics, we numerically integrate the Hamiltonian geodesic equations in the self-dual spacetime and compare the resulting trajectories with the classical Schwarzschild case. In particular, we focus on the dependence of the orbital structure on the polymeric parameter $P$ and the minimal area parameter $a_0$.

\begin{figure}[htbp]
    \centering
    \includegraphics[width=0.45\textwidth]{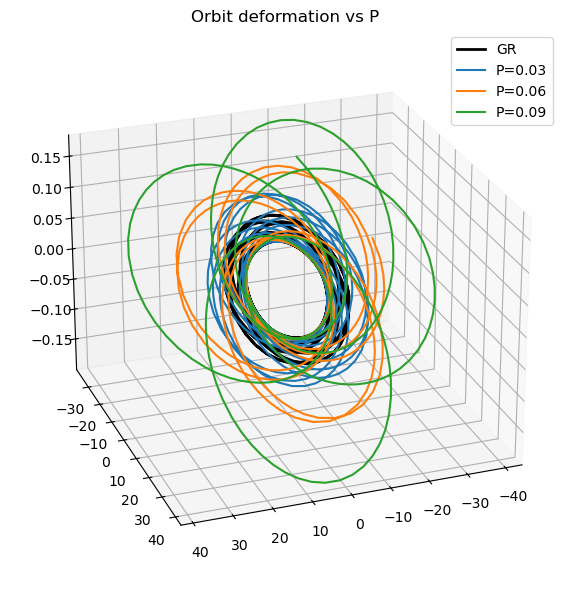}
    \caption{Orbital deformation as a function of the polymeric parameter $P$. The black curve represents the GR case, while colored curves correspond to different values of $P$. The minimal area parameter is fixed at $a_0 = 0.01$.}
    \label{fig:orbit_P}
\end{figure}

Figure~\ref{fig:orbit_P} shows the orbital deformation induced by varying the polymeric parameter $P$, while keeping $a_0 = 0.01$ fixed. The black curve corresponds to the classical general relativistic (GR) trajectory, whereas the colored curves represent different LQG configurations. We observe that increasing $P$ leads to a gradual deformation of the orbital precession pattern. In particular, the deviation from the GR orbit becomes more pronounced in the strong-field region, indicating that $P$ effectively modifies the near-horizon redshift structure and the effective radial potential.

\begin{figure}[htbp]
    \centering
    \includegraphics[width=0.45\textwidth]{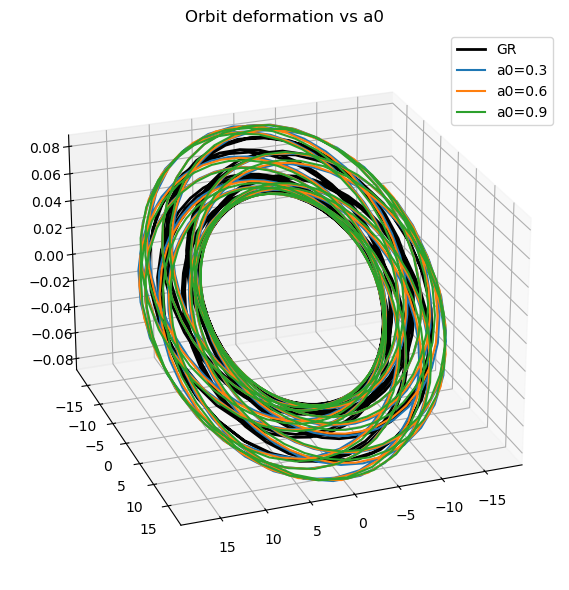}
    \caption{Orbital deformation as a function of the minimal area parameter $a_0$. The black curve represents the GR case, while colored curves correspond to different values of $a_0$. The polymeric parameter is fixed at $P = 0.01$.}
    \label{fig:orbit_a0}
\end{figure}

On the other hand, Figure~\ref{fig:orbit_a0} illustrates the effect of the minimal area parameter $a_0$ while keeping $P = 0.01$ fixed. Unlike $P$, which affects the global orbital structure, the parameter $a_0$ primarily influences the small-radius regime. As $a_0$ increases, the orbit exhibits a noticeable smoothing near the central region, reflecting the regularization of the classical singularity induced by quantum geometry effects. This leads to a reduction in orbital compression near periapsis and slightly modifies the precession rate.

Overall, these results demonstrate that LQG corrections introduce two distinct effects on orbital dynamics: $P$ controls the global deformation of the trajectory, while $a_0$ governs the ultraviolet regularization of the central region.

\subsection{Gravitational waveform and LQG-induced modifications}

To assess the observational impact of loop quantum gravity (LQG) corrections, we compute the leading-order gravitational waveform generated by the orbital motion using the standard quadrupole approximation. In this framework, the dominant $h_{+}$ polarization is determined by the second time derivative of the mass quadrupole moment of the system,
\begin{equation}
h_{+}(t) \simeq \frac{2}{r_{\rm obs}} \ddot{Q}(t),
\end{equation}
where $r_{\rm obs}$ denotes the distance to the observer and $Q$ is constructed from the particle trajectory.

\begin{figure}[htbp]
    \centering
    \includegraphics[width=0.45\textwidth]{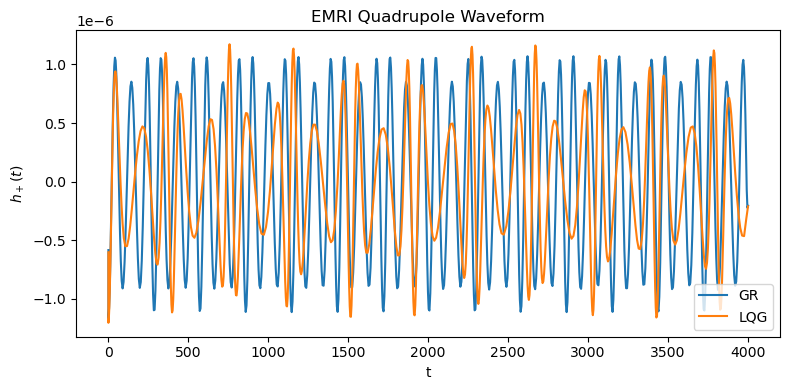}
   \caption{
Quadrupole gravitational waveform $h_{+}(t)$ generated by test-particle geodesic motion in the Schwarzschild (GR) and self-dual LQG spacetimes.
The GR waveform is shown in blue, while the LQG-corrected waveform is shown in orange.
Under identical initial conditions, the two signals remain nearly indistinguishable at early times but gradually deviate during the evolution due to the modification of the orbital dynamics induced by quantum-gravity corrections.
These waveform differences reflect the influence of LQG corrections on the strong-field geodesic motion and may provide potential observational signatures for EMRI systems.
}
    \label{fig:waveform}
\end{figure}

Figure~\ref{fig:waveform} shows the resulting waveforms for both the Schwarzschild (GR) case and the self-dual LQG-corrected spacetime under identical initial conditions. Although the instantaneous amplitudes remain comparable, the LQG correction leads to a gradual deviation from the GR waveform during the evolution. This difference originates from the modification of the effective radial potential, which changes the orbital dynamics and consequently affects the emitted gravitational-wave signal over long timescales.

To investigate the impact of loop quantum gravity (LQG) corrections on gravitational-wave emission, we compute the quadrupole waveform $h_+(t)$ generated by a test particle moving in the self-dual spacetime. The waveform is obtained from the numerically integrated geodesic trajectories and evaluated using the standard quadrupole approximation.

\begin{figure}[htbp]
    \centering
    \includegraphics[width=0.45\textwidth]{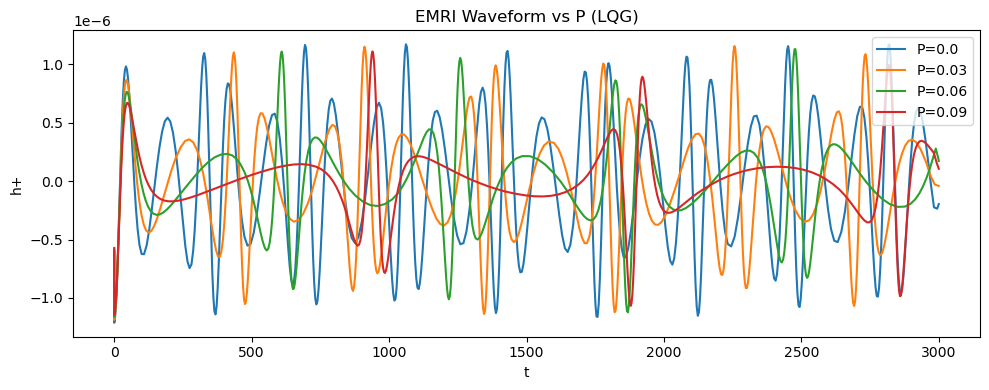}
    \caption{
Gravitational-wave strain $h_+(t)$ for different values of the polymeric parameter $P$, with fixed $a_0=0.01$.
The variation of $P$ induces systematic deviations in the waveform evolution, reflecting the modifications of the orbital dynamics in the LQG-corrected spacetime.
}
    \label{fig:waveform_P}
\end{figure}

Figure~\ref{fig:waveform_P} presents the gravitational-wave signals for different values of the polymeric parameter $P$, while keeping the minimal area parameter fixed at $a_0=0.01$. We observe that increasing $P$ leads to a gradual deviation of the waveform from the classical case during long-time evolution. This difference originates from the modification of the effective radial potential and the resulting changes in the orbital dynamics. Although the instantaneous amplitudes remain comparable, the accumulated waveform deviation becomes more pronounced, indicating that $P$ affects the secular evolution of the orbital motion.

\begin{figure}[htbp]
    \centering
    \includegraphics[width=0.45\textwidth]{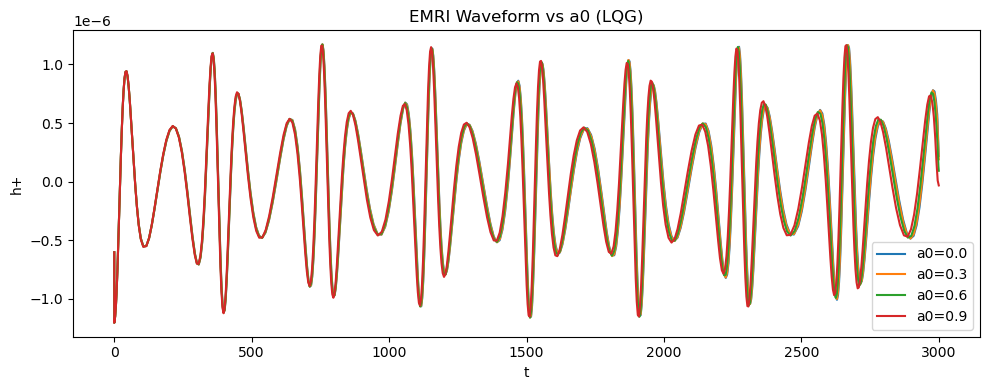}
   \caption{
Gravitational-wave strain $h_+(t)$ for different values of the minimal area parameter $a_0$, with fixed $P=0.01$.
The parameter $a_0$ mainly affects the strong-field orbital dynamics, leading to localized deviations in the waveform evolution during the late stage.
}
    \label{fig:waveform_a0}
\end{figure}

Figure~\ref{fig:waveform_a0} shows the dependence of the waveform on the minimal area parameter $a_0$, with fixed $P=0.01$. In contrast to $P$, the parameter $a_0$ mainly affects the small-radius regime of the orbital dynamics. Therefore, its influence on the waveform is more localized and appears as subtle deviations during the late-time evolution rather than a global modification of the waveform. Larger values of $a_0$ regularize the near-singularity region and modify the strong-field orbital behavior, leading to corresponding changes in the emitted gravitational-wave signal.

Overall, the results indicate that the polymeric parameter $P$ mainly affects the long-term orbital evolution, while the minimal area parameter $a_0$ primarily modifies the strong-field regime near the central black hole. These two parameters therefore introduce different types of modifications in the gravitational-wave signal, which may provide potential observational signatures for future space-based detectors such as LISA.

\section{Phase-cycle structure and waveform locking}

In previous sections, we analyzed the effects of LQG deformation parameters on the global orbital morphology and waveform evolution. In this section, we focus on a more fundamental structural feature: the intrinsic correspondence between orbital phase evolution and the emitted quadrupolar signal.

Rather than emphasizing parameter dependence, we adopt a phase-resolved description in which the waveform is organized according to the orbital angular phase.

\subsection{Phase-cycle decomposition of the orbit}

In the numerical evolution, we use the accumulated azimuthal angle $\phi(t)$ as a natural phase variable and partition the orbital motion into $2\pi$ cycles:
\begin{equation}
\Delta \phi \in [2\pi n, 2\pi(n+1)], \quad n \in \mathbb{N}.
\label{eq:phase_cycle}
\end{equation}

Each $2\pi$ increment corresponds to one complete orbital revolution. Therefore, the trajectory can be decomposed into a sequence of phase-labeled segments, rather than time-based intervals.

The orbital motion is then represented as
\begin{equation}
\mathcal{O}(t) = \bigcup_{n} \mathcal{O}_n(t),
\quad \text{with } \mathcal{O}_n \equiv \{t \,|\, \phi(t)\in [2\pi n,2\pi(n+1)]\}.
\label{eq:orbit_decomposition}
\end{equation}

As shown in Fig.~\ref{fig:P_orbit_waveform} and Fig.~\ref{fig:a0_orbit_waveform}, this phase-based decomposition robustly captures the repeating geometric structure of the orbit under different LQG deformations.

\subsection{Phase-locked structure of the waveform}

More importantly, the quadrupolar waveform $h(t)$ exhibits a one-to-one correspondence with orbital phase cycles. We define the waveform decomposition as
\begin{equation}
h(t) = \sum_{n} h_n(t),
\quad \text{with } t \in \mathcal{O}_n.
\label{eq:waveform_decomposition}
\end{equation}

This implies that each orbital revolution corresponds to a quasi-periodic waveform segment in the gravitational signal.

\begin{figure}[t]
    \centering
    \includegraphics[width=0.48\textwidth]{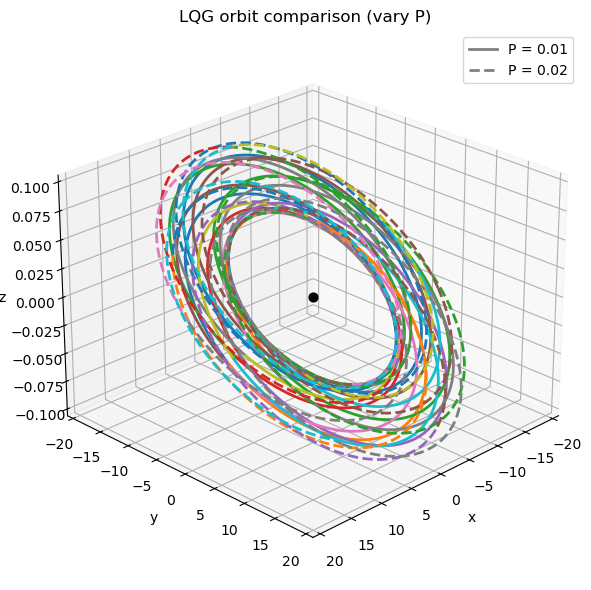}
    \includegraphics[width=0.48\textwidth]{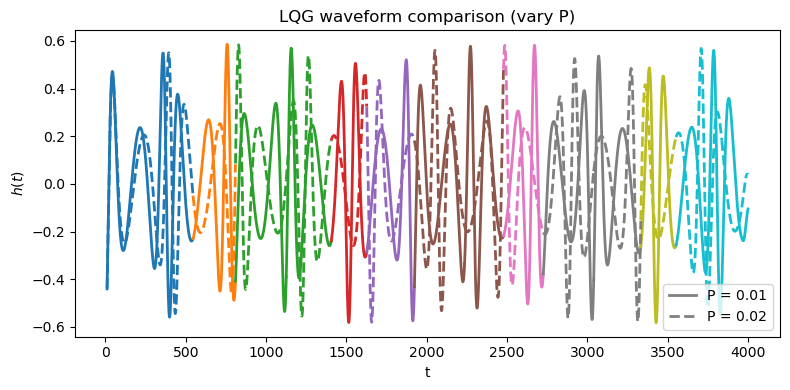}
    \caption{Orbital trajectories and waveform for varying $P$. Solid curves correspond to $P=0.01$, dashed curves correspond to $P=0.02$.}
    \label{fig:P_orbit_waveform}
\end{figure}

\begin{figure}[t]
    \centering
    \includegraphics[width=0.48\textwidth]{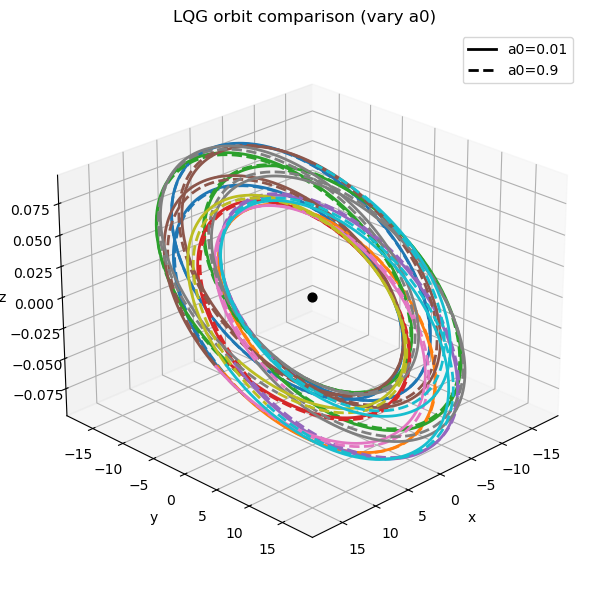}
    \includegraphics[width=0.48\textwidth]{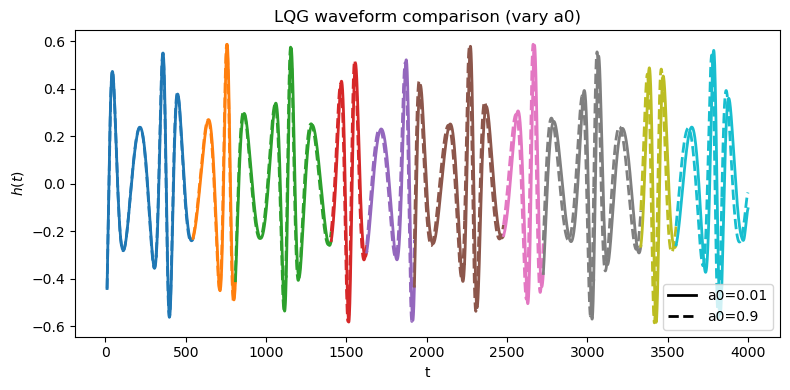}
    \caption{Orbital trajectories and waveform for varying $a_0$. Solid curves correspond to $a_0=0.01$, dashed curves correspond to $a_0=0.9$.}
    \label{fig:a0_orbit_waveform}
\end{figure}

Figures~\ref{fig:P_orbit_waveform}--\ref{fig:a0_orbit_waveform} show that each waveform segment is strictly synchronized with the corresponding orbital cycle, forming a phase-locked structure between orbital motion and radiation.

This structure is characterized by two key properties:
\begin{itemize}
    \item Cycle boundaries are determined by the angular phase $\phi$, rather than coordinate time $t$;
    \item Each waveform segment preserves a self-similar oscillatory structure, while allowing slow secular phase evolution.
\end{itemize}

\subsection{Physical interpretation}

The results indicate that the gravitational-wave signal is not merely a time series, but a phase-structured observable governed by orbital dynamics. In this picture, the orbital phase acts as a natural clock, while the waveform provides a phase-resolved readout.

Consequently, the effects of LQG corrections can be interpreted as:
\begin{equation}
\Delta \Phi_{\text{GW}} = \Delta \Phi(P) + \Delta \Phi(a_0),
\label{eq:phase_shift}
\end{equation}
where different deformation parameters contribute differently to the accumulated phase evolution.

In particular, the parameter $P$ induces a secular phase drift, while $a_0$ primarily modifies the local intra-cycle structure.

This phase-cycle locking structure provides a refined framework for analyzing gravitational-wave signals in modified gravity spacetimes and may be particularly relevant for long-duration observations such as EMRIs.

\section{ LQG Kerr Geodesic Motion and Gravitational-Wave Signatures}

\subsection{Metric and Hamiltonian formulation}

To construct the rotating extension of the self-dual loop quantum gravity (LQG) black hole, we employ the revised Newman--Janis algorithm (NJA). Compared with the original complexification procedure, the revised NJA provides a systematic way to generate an axisymmetric spacetime from the static seed metric while preserving the correct asymptotic behavior.

The resulting rotating LQG-corrected metric can be written as
\begin{equation}
\begin{aligned}
ds^2={}&
-\left(1-\frac{2mr}{\rho^2}\right)dt^2
-\frac{4amr\sin^2\theta}{\rho^2}dt\,d\phi\\
&+\frac{\rho^2}{\Delta}dr^2
+\rho^2d\theta^2
+\frac{\Sigma\sin^2\theta}{\rho^2}d\phi^2 ,
\end{aligned}
\end{equation}
where the metric functions are defined by
\begin{equation}
\rho^2=K(r)+a^2\cos^2\theta ,
\end{equation}

\begin{equation}
\Delta=F(r)r^2+a^2 ,
\end{equation}

\begin{equation}
\Sigma=
\left(K(r)+a^2\right)^2
-a^2\Delta\sin^2\theta ,
\end{equation}

with the radial deformation function
\begin{equation}
K(r)=r^2\sqrt{\frac{F(r)}{G(r)}} .
\end{equation}

Here, $a=J/M$ denotes the spin parameter, while $m$ represents the ADM mass of the rotating spacetime. In the asymptotic limit $r\rightarrow\infty$, the metric approaches the Kerr geometry and satisfies
\begin{equation}
M_{\rm ADM}
=
\lim_{r\rightarrow\infty}
\frac{r}{2}
\left(F^{-1}(r)-1\right)
=m ,
\end{equation}
ensuring the consistency of the mass parameter with the asymptotic gravitational field.

The rotating LQG spacetime reduces to the Kerr black hole in the classical limit,
\begin{equation}
P\rightarrow0,\qquad a_0\rightarrow0 ,
\end{equation}
while the non-rotating limit,
\begin{equation}
a\rightarrow0 ,
\end{equation}
recovers the static self-dual LQG black hole solution.

This rotating LQG-corrected geometry provides a suitable framework for investigating quantum-gravity effects in rotating compact objects. In particular, the parameters $P$ and $a_0$ modify the near-horizon structure and may leave characteristic imprints on particle dynamics, gravitational-wave emission, and black-hole spectroscopy.

\subsection{Four-velocity normalization}

To ensure physically consistent timelike geodesics, we impose the normalization condition
\begin{equation}
g^{\mu\nu} p_\mu p_\nu = -1,
\end{equation}
which is equivalent to
\begin{equation}
H = -\frac{1}{2}.
\end{equation}

This condition is used to determine the initial energy component $p_t$, ensuring that all trajectories satisfy the proper time parametrization.

\subsection{Initial conditions and numerical setup}

We consider initial conditions
\begin{equation}
r_0 = 10M, \quad \theta_0 = \frac{\pi}{2}, \quad \phi_0 = 0,
\end{equation}
with
\begin{equation}
p_r = 0, \quad p_\theta = 0.02, \quad L_z = p_\phi = 3.8.
\end{equation}

The equations of motion are integrated using a high-precision adaptive solver with uniform sampling to ensure smooth trajectories.

\subsection{Orbital motion}

The spatial trajectory is reconstructed via
\begin{equation}
R = \sqrt{r^2 + \frac{a_0^2}{r^2}},
\end{equation}
and
\begin{equation}
x = R \sin\theta \cos\phi, \quad
y = R \sin\theta \sin\phi, \quad
z = R \cos\theta.
\end{equation}

\begin{figure}[htbp]
\centering
\includegraphics[width=0.45\textwidth]{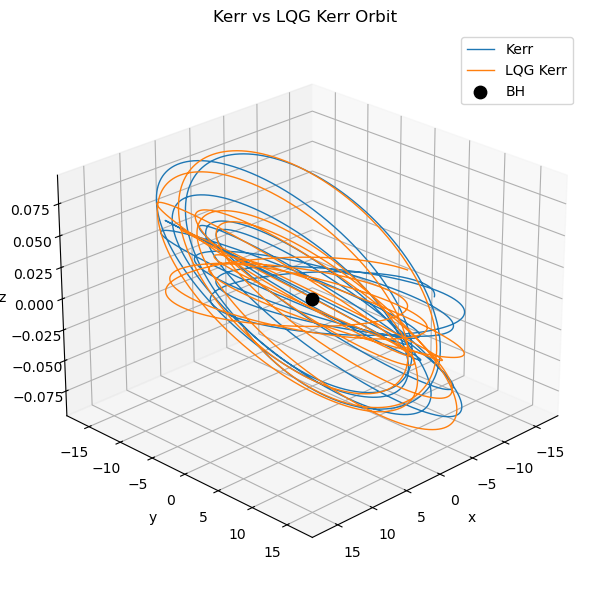}
\caption{Comparison of geodesic motion in Kerr and LQG-corrected rotating spacetime. The central black hole is marked at the origin.}
\label{fig:orbit}
\end{figure}

Figure~\ref{fig:orbit} compares the orbital motion in Kerr and LQG-corrected spacetimes.

\subsection{Quadrupole gravitational-wave signal}

The gravitational-wave signal is computed using the quadrupole approximation:
\begin{equation}
Q_{ij} = \mu \left(x_i x_j - \frac{1}{3} r^2 \delta_{ij}\right),
\end{equation}
and the plus polarization is given by
\begin{equation}
h_+(t) = \frac{2}{D} \frac{d^2 Q}{dt^2},
\end{equation}
where $D$ is the distance to the observer.

\begin{figure}[htbp]
\centering
\includegraphics[width=0.45\textwidth]{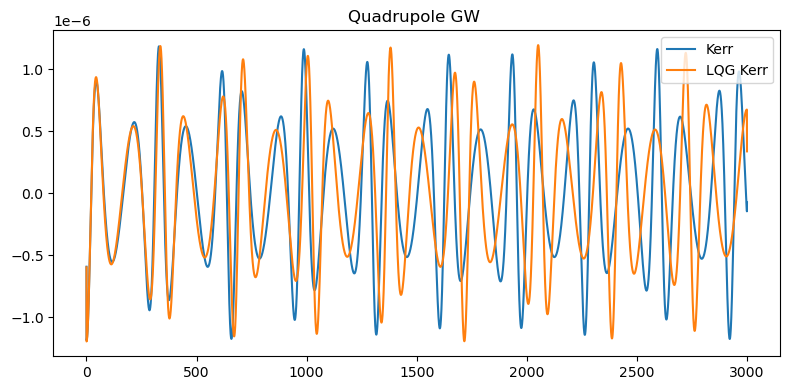}
\caption{Quadrupole gravitational-wave signal comparison between Kerr and LQG-corrected spacetime.}
\label{fig:gw}
\end{figure}

Figure~\ref{fig:gw} shows the comparison between Kerr and LQG-corrected gravitational-wave signals.

The parameter $P$ controls the strength of quantum geometry corrections, while $a_0$ introduces a regularization scale in the radial structure. The results indicate that both orbital dynamics and the corresponding gravitational-wave signals are sensitive to these parameters in the strong-field regime.

\subsection{Inspiral trajectories and gravitational waveforms in LQG and rotating LQG spacetimes.}

We study the dynamics of a test particle in a loop quantum gravity (LQG)-corrected rotating black hole spacetime, focusing on the influence of the quantum deformation parameters and gravitational-wave radiation reaction (RR) on the orbital evolution.

The particle motion is formulated within the Hamiltonian framework,
\begin{align}
\mathcal{H}
=
\frac{1}{2}g^{\mu\nu}p_\mu p_\nu ,
\end{align}
where $g^{\mu\nu}$ is the inverse metric of the rotating LQG spacetime and $p_\mu$ denotes the canonical momentum. The equations of motion are obtained from Hamilton's equations,
\begin{align}
\frac{dx^\mu}{d\tau}
&=
\frac{\partial\mathcal{H}}{\partial p_\mu},
\\
\frac{dp_\mu}{d\tau}
&=
-\frac{\partial\mathcal{H}}{\partial x^\mu}.
\end{align}

To incorporate the dissipative effect associated with gravitational-wave emission, we introduce a phenomenological leading-order quadrupole radiation-reaction prescription. The secular evolution of the orbital energy and angular momentum is described by
\begin{align}
\frac{dE}{d\tau}
&=
-\mathcal{P}_{\rm GW}(r),
\\
\frac{dL_z}{d\tau}
&=
-\mathcal{L}_{\rm GW}(r),
\end{align}
where the gravitational-wave energy and angular momentum fluxes are approximated as
\begin{align}
\mathcal{P}_{\rm GW}(r)
&=
\frac{32}{5}\mu^2 r^4\Omega^6,
\\
\mathcal{L}_{\rm GW}(r)
&=
\frac{32}{5}\mu^2 r^4\Omega^5 .
\end{align}

Here, $\Omega$ represents the orbital angular frequency determined by the circular-orbit dynamics in the LQG-corrected rotating spacetime. The radiation-reaction terms gradually decrease the orbital energy and angular momentum, driving the compact object to inspiral and allowing us to investigate the impact of quantum corrections on the orbital evolution and gravitational-wave emission.

The resulting dynamical system is integrated numerically using a high-precision Runge--Kutta solver with relative and absolute tolerances set to 
\[
\mathrm{rtol}=\mathrm{atol}=10^{-9}.
\]
The initial conditions are chosen as
\begin{align}
r_0 &= 20, \quad 
\theta_0 = \frac{\pi}{2}, \quad 
\phi_0 = 0, \nonumber\\
p_r &= 0, \quad 
p_\theta = 0.02, \quad 
E_0 = 0.955, \quad 
L_{z0}=3.9 .
\end{align}

To visualize the orbital geometry in the LQG-corrected spacetime, we introduce the effective radial coordinate determined by the angular metric function,
\begin{align}
R=\sqrt{H(r)}
=
\sqrt{r^2+\frac{a_0^2}{r^2}},
\end{align}
and transform the trajectory into Cartesian coordinates through
\begin{align}
x &= R\sin\theta\cos\phi, \\
y &= R\sin\theta\sin\phi, \\
z &= R\cos\theta .
\end{align}

We perform a systematic parameter scan over the quantum deformation parameter $P$ to investigate its influence on the orbital morphology and inspiral dynamics. Both the conservative geodesic evolution and the dissipative inspiral evolution driven by gravitational-wave radiation reaction are considered.

\begin{figure*}[htbp]
    \centering
    \includegraphics[width=0.85\textwidth]{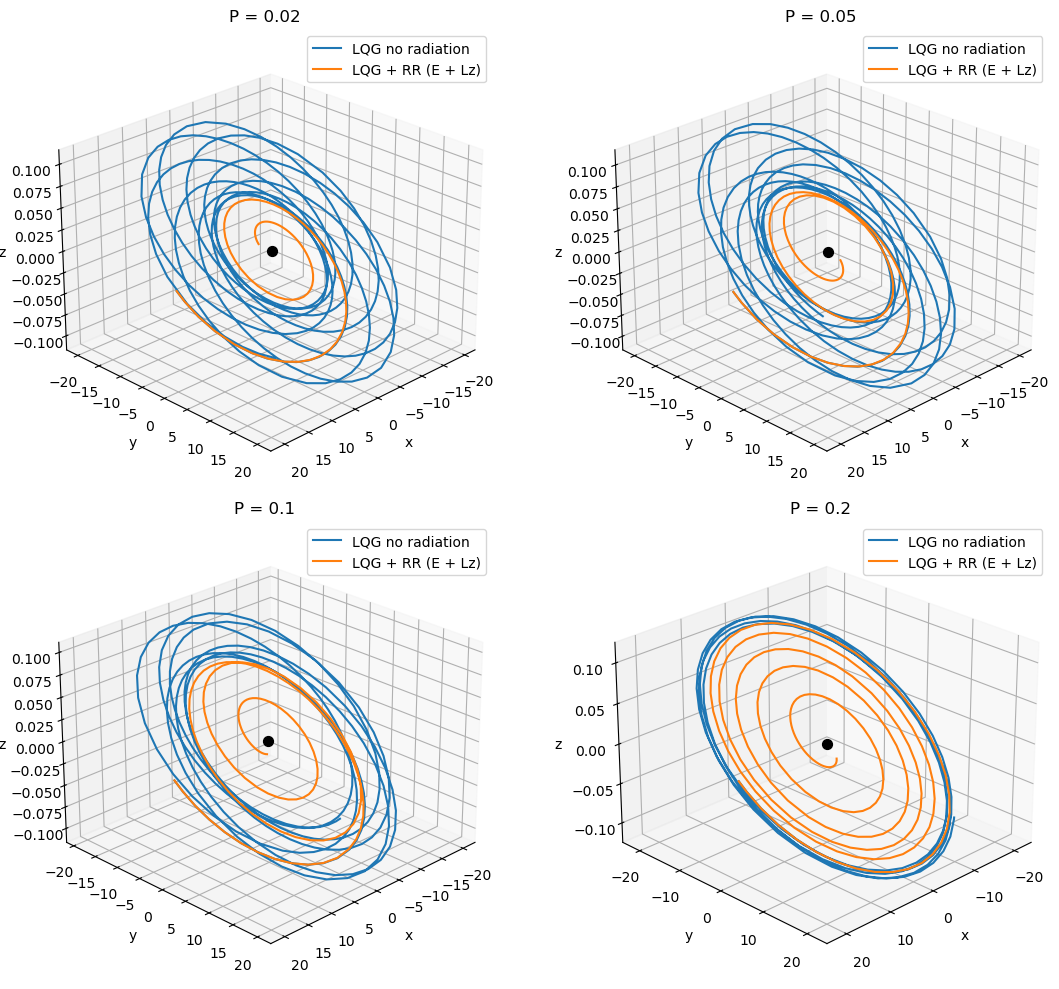}
    \caption{Orbital dynamics in the LQG-corrected rotating spacetime for different values of the deformation parameter $P$.
Each panel corresponds to a different $P$ value, comparing conservative geodesic motion (blue) and dissipative evolution including gravitational-wave radiation reaction (orange).
The central black dot represents the black hole singularity.}
   \label{fig:LQG_orbits_P}
\end{figure*}

Figure~\ref{fig:LQG_orbits_P} presents the resulting trajectories for different values of $P$. Each panel compares conservative and dissipative motion. We observe that increasing $P$ leads to stronger deformation of the orbital morphology in the strong-field regime. When radiation reaction is included, the orbit exhibits a gradual inspiral driven by secular losses of energy and angular momentum. The combined effect of LQG corrections and gravitational-wave backreaction induces noticeable modifications in the long-term orbital dynamics, which may leave imprints on EMRI gravitational-wave signals.

\begin{figure*}[htbp]
    \centering
     \includegraphics[width=0.85\textwidth]{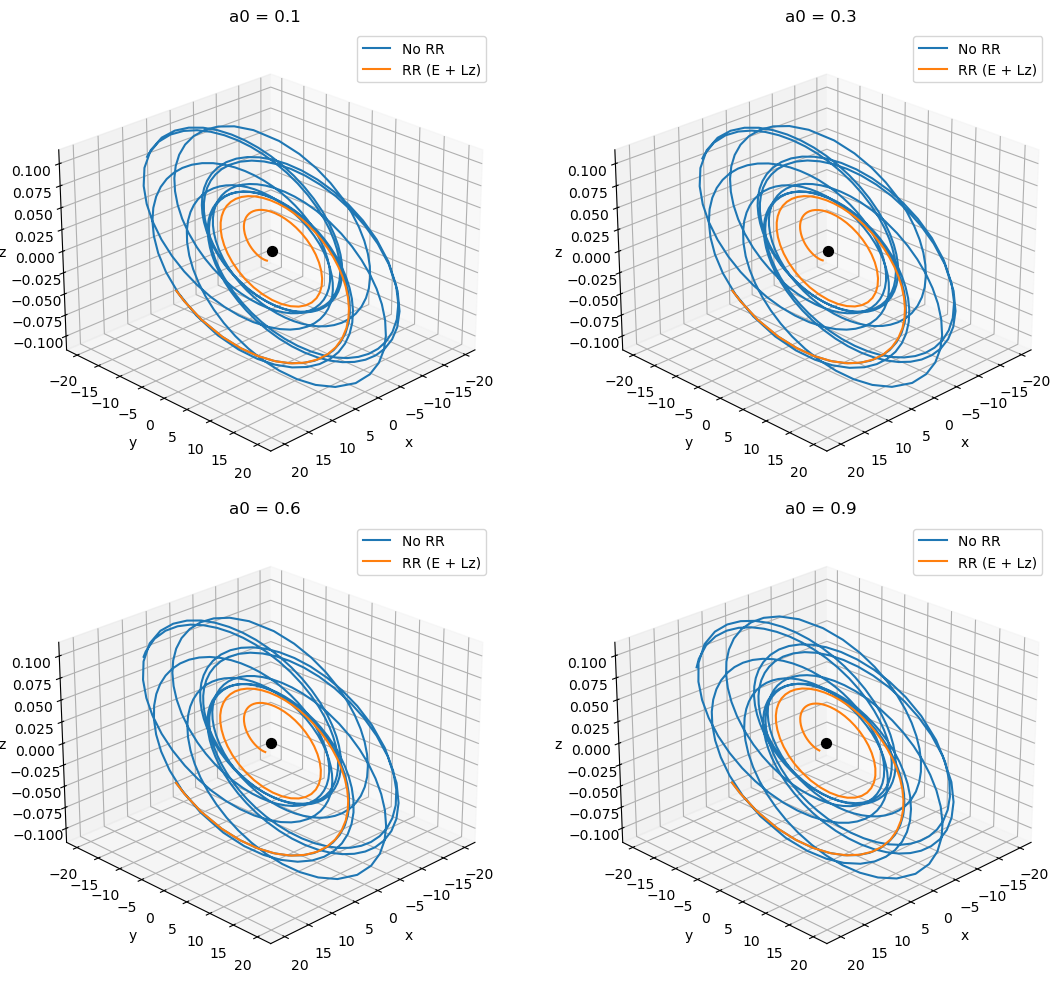}
    \caption{
Orbital dynamics for different values of the quantum geometry parameter $a_0$ with fixed $P=0.03$.
Each panel compares conservative geodesic motion (blue) and dissipative evolution including gravitational-wave radiation reaction (orange).
The effect of $a_0$ is found to be subdominant compared with the deformation parameter $P$, leading only to mild corrections in the orbital morphology.
}
    \label{fig:a0_scan}
\end{figure*}

To further assess the role of the quantum geometry scale, we fix the deformation parameter to $P=0.03$ and vary the parameter $a_0$. The corresponding orbital evolution is shown in Fig.~\ref{fig:a0_scan}. 

We observe that, in contrast to the significant impact of $P$, the parameter $a_0$ produces only a mild modification of the orbital morphology. The overall qualitative structure of the trajectories remains unchanged, both in the conservative and radiation-reaction-included cases. The main effect of increasing $a_0$ is a modification of the angular sector of the metric through $H(r)$, which induces weak corrections to the orbital precession while leaving the global inspiral behavior unchanged.

\begin{figure*}[htbp]
  \centering
  \includegraphics[width=0.85\textwidth]{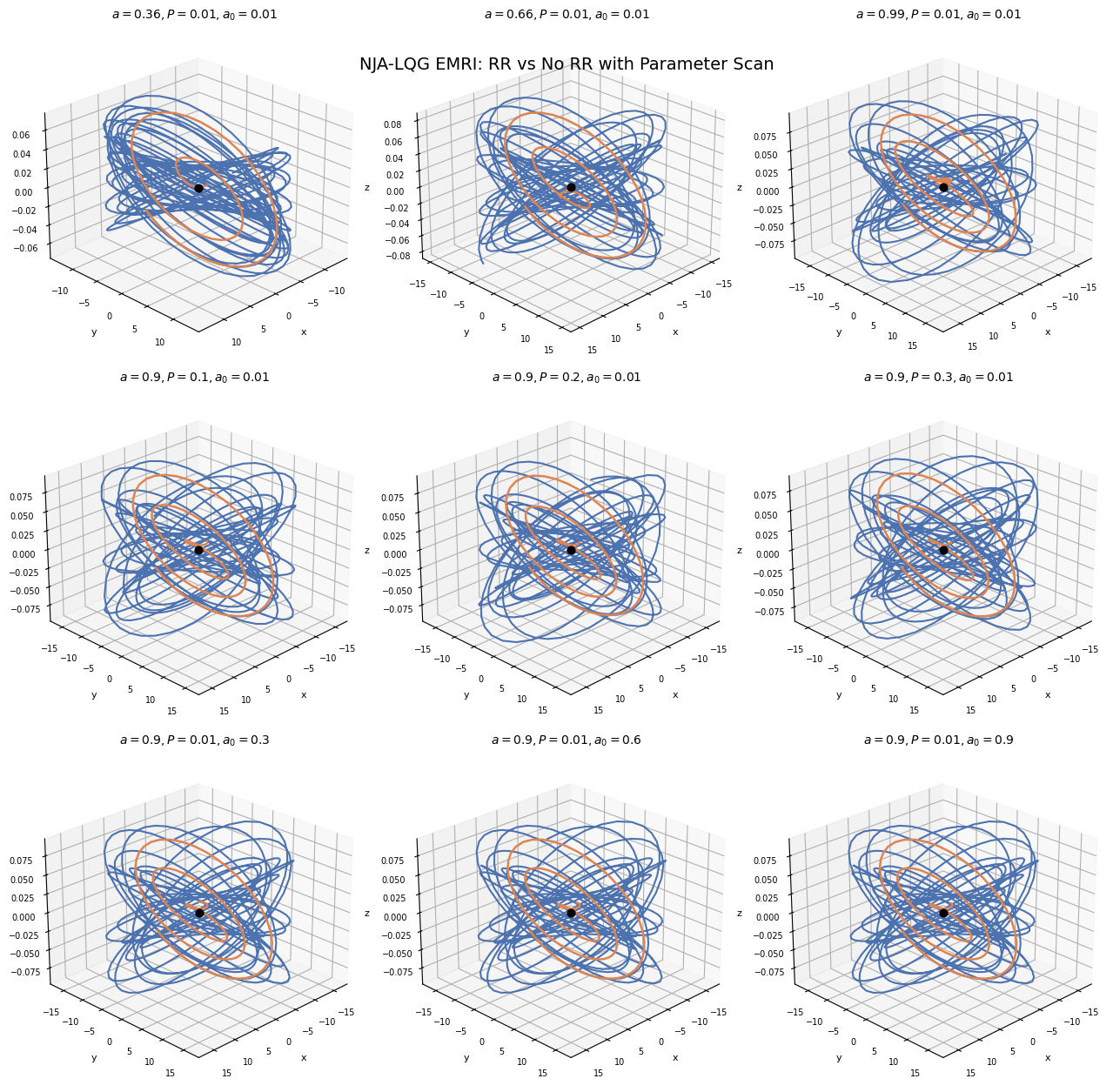}
 \caption{
Three-by-three parameter scan of EMRI trajectories in the Newman–Janis algorithm inspired LQG (NJLQG) spacetime. 
Each panel compares geodesic motion (No RR, blue curves) with radiation-reaction-driven inspirals (RR, orange curves).
The first row shows variation in the spin parameter $a$ with fixed $P=0.01$ and $a_0=0.01$. 
The second row shows variation in $P$ with fixed $a=0.9$ and $a_0=0.01$. 
The third row shows variation in $a_0$ with fixed $a=0.9$ and $P=0.01$.
We find that the spin parameter $a$ has the dominant effect on orbital dynamics, while $P$ induces weaker corrections and $a_0$ has the smallest influence.
}
   \label{fig:9panel}
\end{figure*}

In Fig.~\ref{fig:9panel}, we present a systematic $3 \times 3$ parameter scan of EMRI trajectories in the NJA-LQG spacetime. Each panel compares geodesic motion without radiation reaction (No RR, blue curves) and dissipative inspiral trajectories including radiation reaction (RR, orange curves), allowing a direct visualization of both conservative and dissipative effects.

The first row shows the dependence on the Kerr spin parameter $a$, while keeping $P=0.01$ and $a_0=0.01$ fixed. As $a$ increases, the orbital structure exhibits increasingly strong frame-dragging effects, leading to more pronounced precession and deformation of the orbital geometry. This indicates that the spin parameter plays the dominant role in determining the qualitative morphology of the trajectories.

The second row explores the effect of the loop quantum gravity correction parameter $P$ with fixed $a=0.9$ and $a_0=0.01$. We observe that variations in $P$ lead to relatively mild modifications of the orbital evolution. Once radiation reaction is included, these deviations become even less pronounced, suggesting that dissipative effects tend to partially suppress the influence of global quantum corrections in the inspiral regime.

The third row investigates the impact of the minimal length scale parameter $a_0$ while fixing $a=0.9$ and $P=0.01$. Within the parameter range considered, changes in $a_0$ produce only minor variations in the orbital configuration, mainly affecting the deep strong-field region without significantly altering the overall inspiral morphology.

Overall, the comparison between the three parameter sets reveals a clear hierarchy of influence on EMRI dynamics:
the spin parameter $a$ dominates the orbital behavior, the LQG correction parameter $P$ induces subleading effects, and the minimal length scale parameter $a_0$ contributes the weakest modifications. This hierarchy is consistent with the expectation that classical relativistic effects dominate over quantum-gravity-induced corrections in the weak-to-intermediate field regime relevant for most of the inspiral evolution.

\subsection{EMRI dynamics and gravitational-wave emission in NJLQG spacetime}

To investigate the dynamical and observational implications of the NJLQG spacetime, we simulate the orbital evolution of an extreme mass-ratio inspiral (EMRI) using the full covariant form of the metric. The geometry is constructed from the LQG-deformed seed functions $F_{\mathrm{LQG}}(r)$ and $G_{\mathrm{LQG}}(r)$, together with the derived quantities $K(r)$, $M_{\mathrm{eff}}(r)$, and $\Delta(r)$, which encode both quantum and rotational corrections.

Radiation reaction effects are included through leading-order energy and angular momentum fluxes, leading to a secular inspiral of the orbit. The particle motion is obtained by integrating the Hamiltonian equations associated with the inverse metric $g^{\mu\nu}$.

\begin{figure}[t]
\centering

\includegraphics[width=0.75\linewidth]{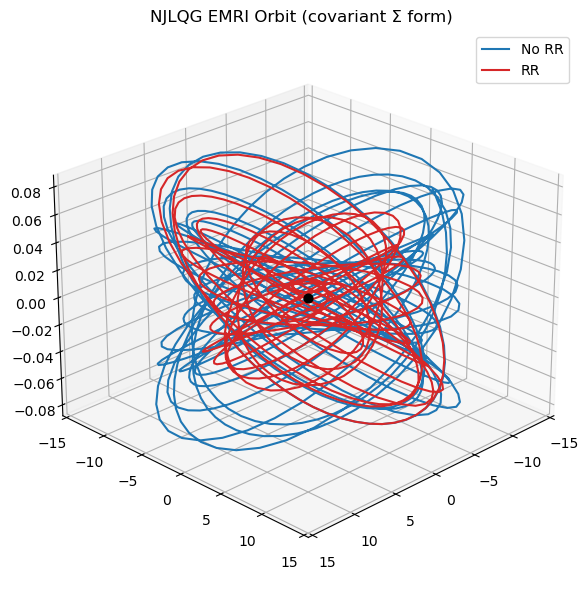}

\vspace{0.3cm}

\includegraphics[width=0.75\linewidth]{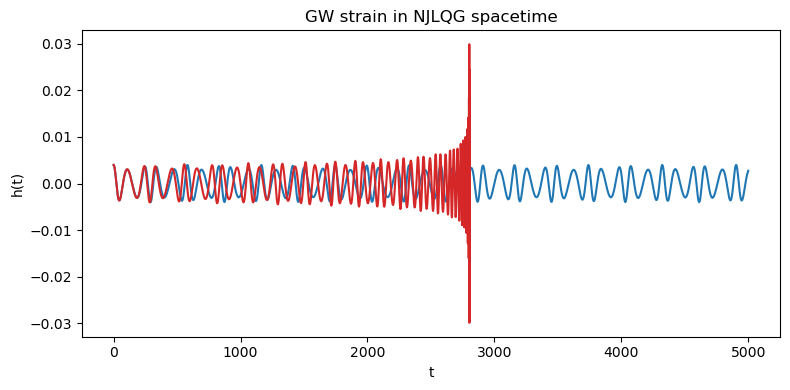}

\caption{
EMRI dynamics in NJLQG spacetime. Top: orbital trajectories with and without radiation reaction. Bottom: gravitational-wave strain $h(t)$. Radiation reaction induces orbital inspiral and noticeable waveform phase evolution.
}
\label{fig:emri_njlqg}
\end{figure}

The resulting orbital dynamics and gravitational-wave signals are shown in Fig.~\ref{fig:emri_njlqg}. The left panel illustrates the orbital trajectories with and without radiation reaction. In the absence of radiation reaction, the orbit remains quasi-periodic and stable. In contrast, the inclusion of dissipative effects induces a gradual inspiral toward the central region, clearly demonstrating orbital decay due to gravitational-wave emission.

Fig.~\ref{fig:emri_njlqg} shows the corresponding gravitational-wave strain $h(t)$ computed from a quadrupole-inspired approximation. The waveform exhibits clear phase evolution differences between the two cases, indicating that radiation reaction and the underlying NJLQG geometry both contribute to observable modifications in the signal. In particular, quantum-gravity-induced deformations affect the orbital precession and amplitude modulation through the modified effective potential structure.

Overall, the results demonstrate that EMRI systems in NJLQG spacetime exhibit distinctive orbital evolution and gravitational-wave signatures, providing a potential observational channel to probe deviations from classical Kerr geometry.

\subsection{Parameter estimation of the LQG–Kerr deformation model}

In order to quantify the measurability of the deformation parameters in the LQG–Kerr spacetime, we construct an effective observable-based Fisher information framework. Instead of relying on quasi-normal modes or full waveform modeling, we define a phenomenological observable vector that captures the dominant dynamical imprints of the geometry on orbital motion.

The rotating LQG spacetime is characterized by three parameters,
\begin{equation}
\theta=(a,P,a_0),
\end{equation}
where $a$ denotes the spin parameter, $P$ represents the polymeric quantum correction, and $a_0$ is the minimal area parameter.

The metric deformation is introduced through the LQG seed functions,
\begin{equation}
G_{\rm LQG}(r)
=
\frac{(r-r_+)(r-r_-) (r+r_\ast)^2}
{r^4+a_0^2},
\end{equation}

and

\begin{equation}
F_{\rm LQG}(r)
=
\frac{(r-r_+)(r-r_-)r^4}
{(r+r_\ast)^2(r^4+a_0^2)},
\end{equation}

where

\begin{equation}
r_+=2M,\qquad
r_-=2MP^2,\qquad
r_\ast=2MP .
\end{equation}

Using the Newman--Janis construction, the rotating geometry is determined by

\begin{equation}
K(r)=r^2
\sqrt{\frac{F_{\rm LQG}(r)}
{G_{\rm LQG}(r)}},
\end{equation}

\begin{equation}
M_{\rm eff}(r)
=
\frac{K(r)-F_{\rm LQG}(r)r^2}
{2r},
\end{equation}

and

\begin{equation}
\Delta(r)=F_{\rm LQG}(r)r^2+a^2 .
\end{equation}

To quantify the sensitivity of orbital dynamics to the LQG parameters, we construct an effective Fisher information analysis based on geometrical observables extracted from the metric.

The observable vector is defined as

\begin{equation}
\mathcal{O}(\theta)
=
\left\{
g_{tt}(r_i),
g_{t\phi}(r_i),
g_{\phi\phi}(r_i),
\Delta(r_i)
\right\}_{i=1}^{N},
\end{equation}

where the metric components are evaluated from the full covariant NJA-LQG metric.

The Fisher matrix is then computed as

\begin{equation}
F_{ij}
=
\frac{1}{\sigma^2}
\sum_{k=1}^{N}
\frac{\partial \mathcal{O}_k}
{\partial\theta_i}
\frac{\partial \mathcal{O}_k}
{\partial\theta_j},
\end{equation}

where $\sigma$ denotes an effective uncertainty parameter.

The covariance matrix is obtained through

\begin{equation}
\Sigma=F^{-1}.
\end{equation}

Assuming a Gaussian approximation around the fiducial parameter values,
\begin{equation}
\theta
\sim
\mathcal{N}(\theta_0,\Sigma),
\end{equation}

we obtain the corresponding parameter constraints and correlations.

The resulting constraints are:
\begin{align}
a &= 0.902^{+0.0042}_{-0.0041}, \\
P &= 0.030^{+0.00015}_{-0.00015}, \\
a_0 &= 21.4^{+23}_{-15}.
\end{align}

The posterior distributions show that the spin parameter $a$ and deformation parameter $P$ are well constrained and mildly correlated, while $a_0$ exhibits a significantly larger uncertainty due to its weak imprint on the effective potential structure.

\begin{figure}[htbp]
\centering
\includegraphics[width=0.45\textwidth]{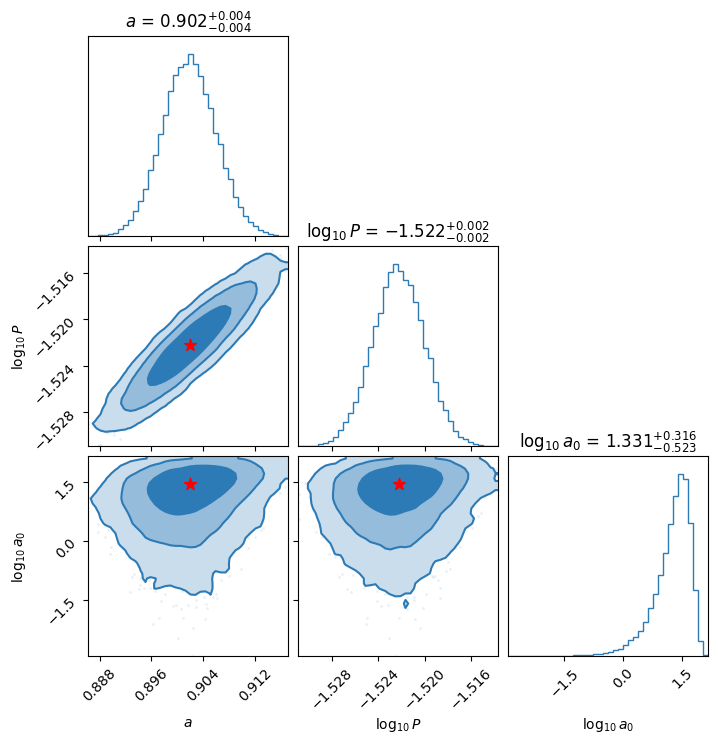}
\caption{
Corner plot of the posterior distribution for the LQG–Kerr deformation parameters $\theta = (a, P, a_0)$ obtained using the Fisher matrix approximation.
The blue contours represent the $68\%$, $95\%$, and $99.7\%$ credible regions.
Red stars indicate the maximum a posteriori (MAP) estimate.
The marginal constraints show that $a$ and $P$ are well constrained, while $a_0$ exhibits large uncertainty, reflecting its weak observational imprint in the effective orbital observables.
}
\label{fig:corner_lqg_kerr}
\end{figure}

\section{Conclusions}

In this work, we have systematically investigated the orbital dynamics and gravitational-wave radiation properties of extreme mass-ratio inspirals (EMRIs) in loop quantum gravity (LQG)-corrected black hole spacetimes. By considering both static and rotating LQG black hole geometries, we have analyzed the effects of quantum correction parameters, including the polymeric parameter $P$ and the minimal area parameter $a_0$, on test-particle geodesic motion, orbital evolution, and gravitational-wave waveforms. Our study establishes a theoretical framework for investigating the dynamics and radiation characteristics of EMRIs in LQG-corrected black hole backgrounds, providing a theoretical basis for exploring quantum corrections in strong gravitational fields.

We first constructed the LQG-corrected black hole spacetime, where quantum modifications are encoded in the metric functions $G(r)$, $F(r)$, and $H(r)$. The classical Schwarzschild limit is recovered when $P\rightarrow0$ and $a_0\rightarrow0$. Through the analysis of the effective potential, we investigated the influence of quantum parameters on particle motion and orbital structures. The results show that the polymeric parameter $P$ mainly affects the global structure of the effective potential, while the minimal area parameter $a_0$ primarily modifies the spacetime properties in the near-horizon region. These scale-dependent quantum corrections lead to modifications of the geodesic dynamics of test particles.

For the static LQG black hole spacetime, we calculated the geodesic trajectories of test particles and studied the influence of quantum parameters on orbital configurations and dynamical evolution. We find that the parameter $P$ produces noticeable modifications to the orbital structures, while $a_0$ mainly affects the motion in the strong-field region. When gravitational radiation is taken into account, these quantum corrections are further reflected in the emitted gravitational-wave signals, leading to distinguishable waveform features for different parameter choices.

We then extended our analysis to rotating LQG black hole spacetimes by employing the revised Newman--Janis algorithm. The resulting rotating metric possesses a Kerr-like structure with modified functions $K(r)$, $M_{\rm eff}(r)$, and $\Delta(r)$. Based on this geometry, we investigated the combined effects of black hole spin and quantum corrections on EMRI orbital dynamics. The parameter analysis indicates that the spin parameter $a$ plays a dominant role in determining the orbital morphology, while the parameters $P$ and $a_0$ introduce additional quantum modifications with different strengths. Furthermore, we calculated the gravitational radiation generated by rotating EMRI systems and analyzed the influence of quantum corrections on the corresponding waveforms.

In addition, we explored the relationship between orbital motion and gravitational-wave radiation. By decomposing the waveform according to the azimuthal variable $\phi(t)$, we found that the orbital periodic structure is closely associated with the radiation waveform, where each orbital cycle corresponds to a waveform segment with similar characteristics. This orbital-cycle-based waveform analysis provides an effective approach for studying the gravitational radiation signatures of different black hole models.

Finally, we employed the Fisher information matrix method to investigate the potential constraints on the LQG parameters. The results show that the polymeric parameter $P$ has a relatively stronger influence on the orbital dynamics and gravitational-wave waveforms, and therefore exhibits better constrainability. In contrast, the minimal area parameter $a_0$ produces weaker effects and is more difficult to constrain. The parameter-space analysis also reveals correlations among different parameters, indicating the necessity of multi-parameter analysis when studying LQG corrections.

In summary, this work investigates the geodesic motion and gravitational radiation characteristics of EMRIs in both static and rotating LQG-corrected black hole spacetimes. We systematically analyze how quantum correction parameters modify orbital dynamics and gravitational-wave waveforms. Our results provide a theoretical basis for studying particle motion and gravitational-wave emission in quantum-corrected black hole backgrounds, and offer further insights into the possible manifestations of quantum gravity effects in astrophysical systems.
\begin{acknowledgements}
The authors gratefully acknowledge support from the project ``Integrated Electronics Technology for Inertial Sensors'' (2024YFC2207003) led by Meilin Liu. 

This work is also supported by the China--Brazil Belt and Road Joint Laboratory on Radio Astronomy Technology, and the National Key Research and Development Program of China under the project ``Strategic Science and Technology Innovation Cooperation'' (No. 2025YFE0212600), which is also led by Meilin Liu.

The authors also thank all colleagues who provided helpful discussions and assistance during this work, and especially Professor Haiguang Xu for his guidance and support.
\end{acknowledgements}

\section*{Funding}
This work is supported by the National Key Research and Development Program of China (Nos. 2024YFC2207003, 2025YFE0212600) and related programs under the China--Brazil Belt and Road Joint Laboratory on Radio Astronomy Technology.

\section*{Data Availability Statement}
This manuscript has no associated data. 
[Author’s comment: Data sharing is not applicable to this article as no datasets were generated or analyzed during the current study.]

\section*{Code Availability Statement}

Code/software will be made available on reasonable request.

[Author’s comment: The code/software generated during and/or analyzed during the current study is available from the corresponding author on reasonable request.]


\begin{thebibliography}{99}

% =========================
% GW / EMRI / LISA
% =========================
\bibitem{Abbott2016}
B.~P.~Abbott et al. (LIGO Scientific Collaboration),
``Observation of Gravitational Waves from a Binary Black Hole Merger,''
Phys. Rev. Lett. \textbf{116}, 061102 (2016).

\bibitem{Abbott2019}
B.~P.~Abbott et al.,
``GWTC-1: A Gravitational-Wave Transient Catalog,''
Phys. Rev. X \textbf{9}, 031040 (2019).

\bibitem{AmaroSeoane2017}
P.~Amaro-Seoane et al.,
``Laser Interferometer Space Antenna,''
arXiv:1702.00786.

\bibitem{BarackCutler2004}
L.~Barack and C.~Cutler,
``LISA capture sources: Approximate waveforms, signal-to-noise ratios, and parameter estimation accuracy,''
Phys. Rev. D \textbf{69}, 082005 (2004).

\bibitem{Gair2013}
J.~R.~Gair, M.~Vallisneri, S.~L.~Larson, and J.~G.~Baker,
``Testing general relativity with low-frequency, space-based gravitational-wave detectors,''
Living Rev. Relativ. \textbf{16}, 7 (2013).

\bibitem{Barack2018}
L.~Barack et al.,
``Black holes, gravitational waves and fundamental physics: a roadmap,''
Class. Quantum Grav. \textbf{35}, 124001 (2018).

\bibitem{Pound2015}
A.~Pound,
``Motion of small objects in curved spacetime: An introduction to gravitational self-force,''
Fund. Theor. Phys. \textbf{179}, 399 (2015).

\bibitem{CutlerFlanagan1994}
C.~Cutler and E.~E.~Flanagan,
``Gravitational waves from merging compact binaries: How accurately can one extract the binary's parameters from the inspiral waveform?''
Phys. Rev. D \textbf{49}, 2658 (1994).

\bibitem{Maselli2020}
A.~Maselli et al.,
``Probing Planckian corrections with EMRIs,''
Phys. Rev. Lett. \textbf{125}, 141101 (2020).

\bibitem{Stein2021}
L.~C.~Stein,
``Extreme mass ratio inspirals as probes of new gravitational physics,''
Phys. Rev. D \textbf{103}, 064033 (2021).

\bibitem{Rovelli2004}
C.~Rovelli,
\textit{Quantum Gravity},
Cambridge University Press (2004).

\bibitem{Ashtekar2004}
A.~Ashtekar and J.~Lewandowski,
``Background independent quantum gravity: a status report,''
Class. Quantum Grav. \textbf{21}, R53 (2004).

\bibitem{Thiemann2007}
T.~Thiemann,
\textit{Modern Canonical Quantum General Relativity},
Cambridge University Press (2007).

\bibitem{Modesto2004}
L. Modesto,
\newblock Disappearance of black hole singularity in quantum gravity,
\newblock \emph{Phys. Rev. D} \textbf{70}, 124009 (2004),
\newblock doi:10.1103/PhysRevD.70.124009,
\newblock arXiv:gr-qc/0407097.

\bibitem{Modesto2006}
L. Modesto,
\newblock Loop quantum black hole,
\newblock \emph{Class. Quantum Grav.} \textbf{23}, 5587 (2006).

\bibitem{SelfDual2009}
L. Modesto and I. Premont-Schwarz,
\newblock Self-dual black holes in loop quantum gravity: Theory and phenomenology,
\newblock \emph{Phys. Rev. D} \textbf{80}, 064041 (2009).

\bibitem{Modesto2010}
L. Modesto,
\newblock Semiclassical loop quantum black hole,
\newblock \emph{Int. J. Theor. Phys.} \textbf{49}, 1649 (2010).

\bibitem{Gambini2013}
R. Gambini and J. Pullin,
\newblock Loop quantization of the Schwarzschild black hole,
\newblock \emph{Phys. Rev. Lett.} \textbf{110}(21), 211301 (2013),
\newblock doi:10.1103/PhysRevLett.110.211301,
\newblock arXiv:1302.5265 [gr-qc].

\bibitem{Thiemann1993}
T. Thiemann and H.A. Kastrup,
\newblock Canonical quantization of spherically symmetric gravity in Ashtekar's selfdual representation,
\newblock \emph{Nucl. Phys. B} \textbf{399}, 211 (1993),
\newblock doi:10.1016/0550-3213(93)90623-W,
\newblock arXiv:gr-qc/9310012.

\bibitem{Campiglia2007}
M. Campiglia, R. Gambini, and J. Pullin,
\newblock Loop quantization of spherically symmetric midi-superspaces,
\newblock \emph{Class. Quantum Gravity} \textbf{24}, 3649 (2007),
\newblock doi:10.1088/0264-9381/24/14/007,
\newblock arXiv:gr-qc/0703135.

\bibitem{Bengtsson1988}
I. Bengtsson,
\newblock Note on Ashtekar's variables in the spherically symmetric case,
\newblock \emph{Class. Quantum Gravity} \textbf{5}, L139 (1988),
\newblock doi:10.1088/0264-9381/5/10/002.

\bibitem{Bojowald2000}
M. Bojowald and H.A. Kastrup,
\newblock Quantum symmetry reduction for diffeomorphism invariant theories of connections,
\newblock \emph{Class. Quantum Gravity} \textbf{17}, 3009 (2000),
\newblock doi:10.1088/0264-9381/17/15/311,
\newblock arXiv:hep-th/9907042.

\bibitem{Bojowald2004}
M. Bojowald and R. Swiderski,
\newblock The volume operator in spherically symmetric quantum geometry,
\newblock \emph{Class. Quantum Gravity} \textbf{21}, 4881 (2004),
\newblock doi:10.1088/0264-9381/21/21/009,
\newblock arXiv:gr-qc/0407018.

\bibitem{Bojowald2005}
M. Bojowald and R. Swiderski,
\newblock Spherically symmetric quantum horizons,
\newblock \emph{Phys. Rev. D} \textbf{71}, 081501 (2005),
\newblock doi:10.1103/PhysRevD.71.081501,
\newblock arXiv:gr-qc/0410147.

\bibitem{Bojowald2006}
M. Bojowald and R. Swiderski,
\newblock Spherically symmetric quantum geometry: Hamiltonian constraint,
\newblock \emph{Class. Quantum Gravity} \textbf{23}, 2129 (2006),
\newblock doi:10.1088/0264-9381/23/6/015,
\newblock arXiv:gr-qc/0511108.

\bibitem{Kuchar1994}
K.V. Kuchar,
\newblock Geometrodynamics of Schwarzschild black holes,
\newblock \emph{Phys. Rev. D} \textbf{50}, 3961 (1994),
\newblock doi:10.1103/PhysRevD.50.3961,
\newblock arXiv:gr-qc/9403003.

\bibitem{Perez2017}
A. Perez,
\newblock Black holes in loop quantum gravity,
\newblock \emph{Rept. Prog. Phys.} \textbf{80}, 126901 (2017).

\bibitem{Bojowald2020}
M. Bojowald,
\newblock Black-hole models in loop quantum gravity,
\newblock \emph{Universe} \textbf{6}, 125 (2020),
\newblock doi:10.3390/universe6080125.

\bibitem{Tu2023}
Z. Tu, H. Zhang, and J. Wu,
\newblock Geodesic motion in self-dual loop quantum gravity black holes,
\newblock \emph{Phys. Rev. D} \textbf{108}, 084048 (2023).

\bibitem{ISCO2024LQG}
L. Zhang, Y. Liu, and Q. Pan,
\newblock ISCO of self-dual LQG black holes,
\newblock \emph{Class. Quantum Gravity} \textbf{41}, 125012 (2024).

\bibitem{SelfDualGeodesics2024}
R. Wang, K. Lin, and J. Jing,
\newblock Geodesic structure in self-dual LQG spacetimes,
\newblock \emph{Eur. Phys. J. C} \textbf{84}, 567 (2024).

\bibitem{ZiKumar2025}
T.~Zi and S.~Kumar,
``Eccentric extreme mass-ratio inspirals: a gateway to probe quantum gravity effects,''
Eur. Phys. J. C \textbf{85}, 592 (2025),
doi:10.1140/epjc/s10052-025-14330-7.

\bibitem{LQGShadow2023}
C. Liu, Z. Zhang, and J. Li,
\newblock Shadow of rotating LQG black holes,
\newblock \emph{Phys. Rev. D} \textbf{107}, 064015 (2023).

\bibitem{QPO2023}
P. Chen, Y. Wang, and S. Chen,
\newblock Quasi-periodic oscillations in LQG black hole spacetimes,
\newblock \emph{Class. Quantum Grav.} \textbf{40}, 185001 (2023).

\bibitem{NewmanJanis1965}
E.~T.~Newman and A.~I.~Janis,
``Note on the Kerr spinning-particle metric,''
J. Math. Phys. \textbf{6}, 915 (1965).

\bibitem{Erbin2017}
H. Erbin,
\newblock The Newman-Janis algorithm and rotating black holes,
\newblock \emph{Universe} \textbf{3}, 13 (2017).

\bibitem{RotatingLQG2024}
X. Zhang, Q. Pan, and J. Jing,
\newblock Rotating LQG black holes and orbital dynamics,
\newblock \emph{Phys. Rev. D} \textbf{110}, 064048 (2024).

\bibitem{Cruz2019}
M.~B.~Cruz, C.~A.~S.~Silva and F.~A.~Brito,
Gravitational axial perturbations and quasinormal modes of loop quantum black holes,
Eur.\ Phys.\ J.\ C \textbf{79}, 157 (2019),
doi:10.1140/epjc/s10052-019-6565-2.

\end{thebibliography}
\end{document}